\documentclass[cits]{PoS}

\PoS{PoS(HEP2005)086}

\newcommand{\xp}{\Xi^{-} \> \pi^{\pm}}
\newcommand{\xbp}{\bar{\Xi}^{+} \> \pi^{\pm}} 
\newcommand{\ksp}{K^{0}_{S}\> p}
\newcommand{\kspb}{K^{0}_{S}\> \bar{p}}
\newcommand{\ksppb}{K^{0}_{S}\> p\>(\bar{p})}
\newcommand{\coll}{Collaboration}
\newcommand{\etal}{ {\it et al.,} }
\newcommand{\gev}{\; \mathrm{GeV}}
\newcommand{\mev}{\; \mathrm{MeV}}
\newcommand{\tp}{\Theta^+}

\title{Results of the searches for pentaquarks with
strangeness in DIS at HERA}

\ShortTitle{Strange pentaquarks in DIS at HERA}

\author{
\speaker{Sergei Chekanov}\thanks{for the H1 and ZEUS Collaborations} \\
HEP division, Argonne National Laboratory,
9700 S.Cass Avenue, Argonne, IL 60439, USA \\ 
E-mail: \email{chekanov@mail.desy.de}}

\abstract{
The $\ksp$ invariant mass
spectrum was reconstructed in several 
kinematic regions with the main emphasis on
the studies of the production mechanism of  the $\tp$ candidate recently
observed by ZEUS. The candidate $\tp$
signal was found to be produced predominantly in the forward hemisphere in the
laboratory frame. This is unlike the case for the $\Lambda(1520)$ or the
$\Lambda_c$, and indicates that the $\tp$ may have an unusual production
mechanism related to proton-remnant fragmentation.
H1 does not observe a signal and sets an  
upper limit at $95\%$ C.L.  which does
not exclude the ZEUS observation.    
}

\FullConference{International Europhysics Conference on High Energy Physics\\
		 July 21st - 27th 2005\\
		 Lisboa, Portugal}

\begin{document}

\section{Introduction}

Recently, ZEUS made an observation~\cite{zeus} of a narrow 
signal in  the vicinity of $1530\mev$
which can be interpreted as a bound state of five quarks, i.e. as a pentaquark, 
$\Theta^+ =uudd\bar{s}$ \cite{zp:a359:305}. The experimental search was performed
using inclusive $ep$ collisions  with the 
proton energy 820/920 GeV and the electron/positron energy of $27.6\gev$.
The data were collected  using  an integrated luminosity of $121$ pb$^{-1}$ 
from HERA~I.

The $\tp$ candidate decaying to $\ksppb$  was  seen in deep inelastic scattering (DIS) 
at $Q^2>20\gev^2$, while no signal was reported in photoproduction ($Q^2\sim 0$).  
This decay channel  is
not exotic since  3-quark states, such as $\Sigma$ baryons,
can also decay to the $\ksppb$.   
Therefore, it is essential to study the kinematics of the observed signal
and to compare it with that for established 3-quark states.  

The absence of a guiding principle of how to compose three quarks to form  a
baryon at the quark-fragmentation stage leads to a few possible
baryon-production mechanisms. In the case of
baryons with five quarks, the situation could be even
more complicated. 
For collisions involving incoming baryon(s), the fragmentation
mechanism could be influenced by a contribution  from the 
proton-remnant system,
thus the baryonic yield in certain kinematic regions of $ep$ collisions is expected be higher 
than in $e^+e^-$ annihilation.  

\section{Evidence for a baryonic state near 1530 decaying to $\ksp$}

$\ksppb$ invariant masses reconstructed in  
photoproduction and DIS using HERA~I data are shown in Fig.~1(left).
The $\ksppb$  distribution has two peaks,
at around $1522\mev$ ($Q^2>20\gev^2$) and near  $2286\mev$.
The first peak is attributed to the $\tp$ candidate state \cite{zeus}.
The second peak corresponds to the established $\Lambda_c$ baryon.
Both peaks are  best seen for $Q^2>20\gev^2$,  i.e. in the region
where the $\Lambda_c$ peak has the largest 
signal-over-background ratio (S/B=0.22).
Since the average charged-track multiplicity 
is $50\%$ higher\footnote{This is due to the fact that photoproduction 
events were taken using a specific trigger requirement 
(jets with $E_T>6-8 \gev$). The hadronic-final state of 
DIS data sample is almost completely trigger-unbiased.} 
for photoproduction than for DIS at $Q^2>1\gev^2$,
the S/B ratio is small for photoproduction.
Therefore, the
significant combinatorial background for the $\ksppb$ spectrum could explain
the non-observation of the $\tp$ candidate for photoproduction~\cite{zeus}.

It is important to note that the decrease of the S/B ratio for high-multiplicity events
exists only for the states whose production is not driven by pure quark fragmentation.
A typical example is  $\Lambda_c$: there could be at most two $\Lambda_c$ per event
independent of the energy contributing to the hadronic-final state. 
The production of charmed quarks is significantly suppressed at the fragmentation stage, 
therefore, a hard QCD subprocess, like $\gamma^{*} g\to c\bar{c}$, is the  necessary
mechanism for the $\Lambda_c$ formation. 
If the $\tp$ production is driven by the diquark system (in which the
diquark is produced by the incoming proton), 
then the $\tp$ peak should also have a small S/B ratio for high multiplicity events.

In contrast, the S/B ratio for $\Lambda(1520)$ reported by ZEUS~\cite{zeus_eps05} is almost the same 
for photoproduction and DIS. This is  a clear indication that $\Lambda(1520)$ is 
solely produced by the quark fragmentation, and,  therefore,  
this baryon  cannot be used as a reference state  for experimental $\tp$ searches. 
 
\begin{center}
\begin{minipage}[c]{0.49\textwidth}
\includegraphics[width=7.5cm,angle=0]{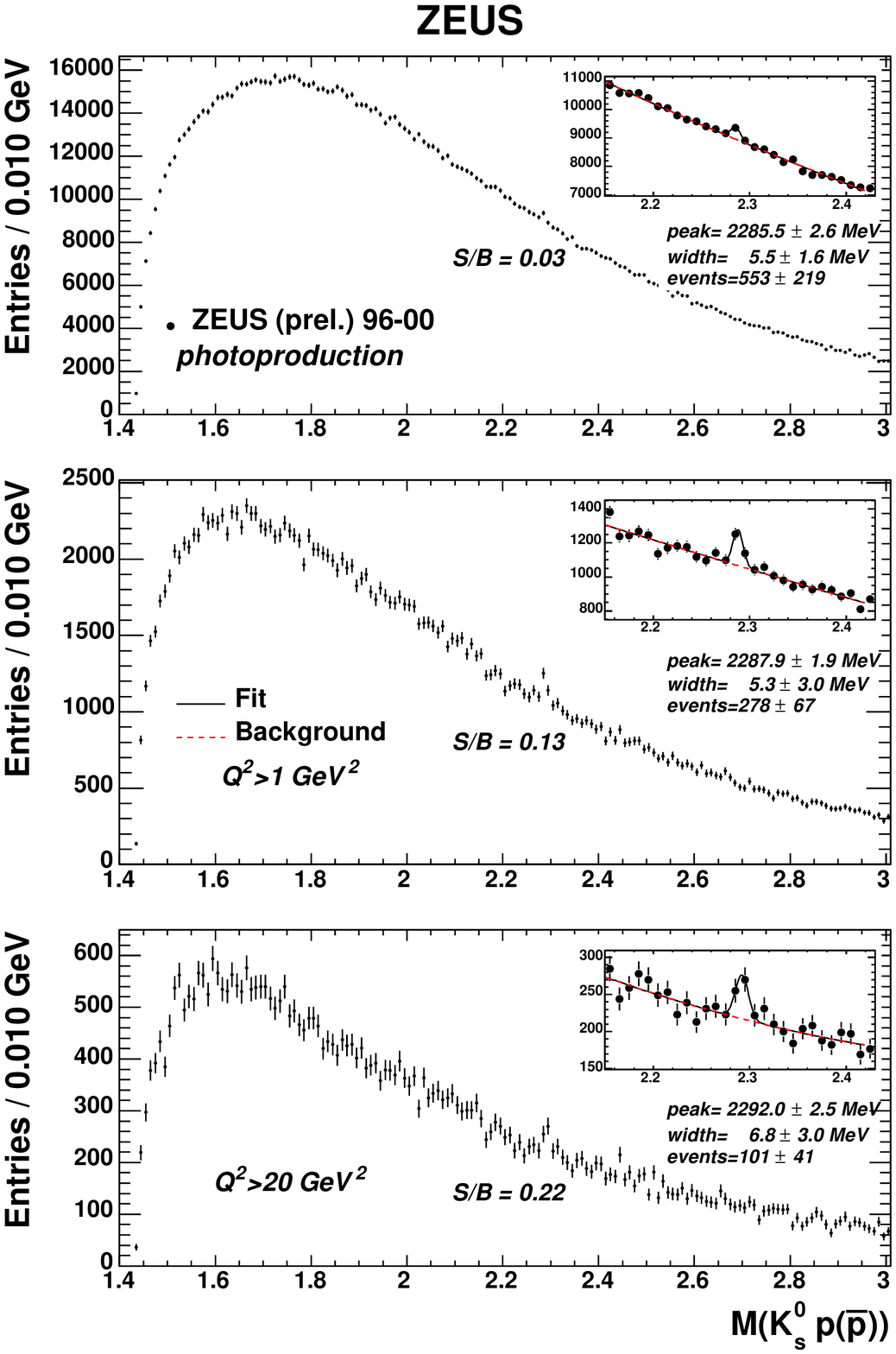}
\label{theta1}
\end{minipage}
\hfill
\begin{minipage}[c]{.49\textwidth}
\includegraphics[width=8.0cm,angle=0]{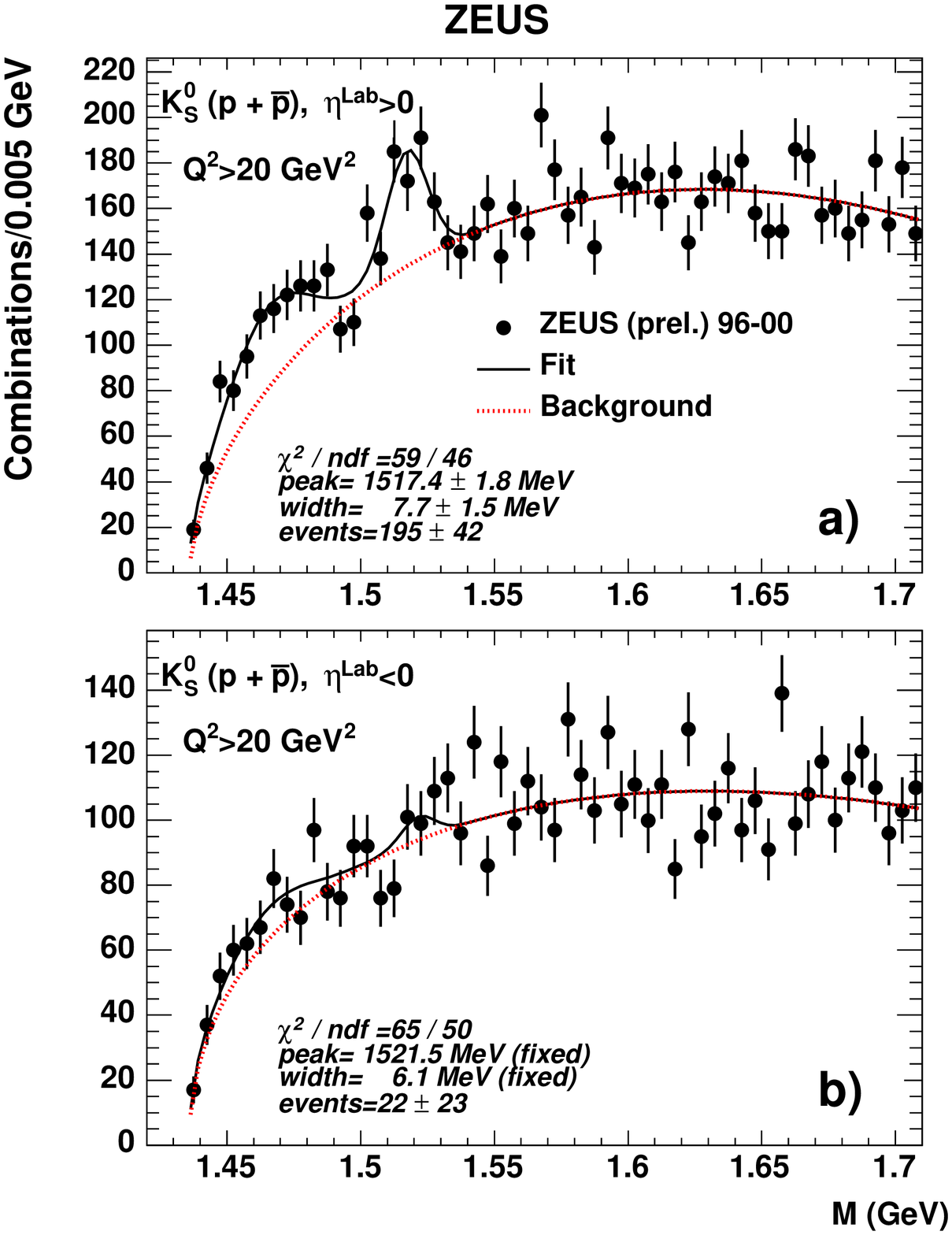}
\label{theta2}
\end{minipage}
\end{center}
\vspace{-0.5cm}
{\bf Figure~1.}~Left: {\it
$\ksppb$ invariant-mass spectra in
photoproduction and DIS  ($Q^{2}>1\gev^{2}$ and $Q^{2}>20\gev^{2}$) with the 
indicated signal-over-background ratio
(S/B) for the $\Lambda_c$ peak.
The insets show the invariant-mass distribution near the $\Lambda_c$ mass
region.} 
Right: 
{\it $\ksppb$ invariant-mass distribution near the  $1520\mev$ mass
for  the forward ($\eta>0$) and rear ($\eta<0$) region.
The fit was performed using a double Gaussian with the threshold function.
}

\vspace{0.5cm}

If the observed $\ksppb$ peak near $1522\mev$ indeed corresponds to a new baryonic state,
then the studies of this peak in different pseudorapidity regions  can help to
understand  the production mechanism of this state.
The  $1522\mev$  peak should be  seen for both forward and rear pseudorapidity regions
if the $\tp$ is produced by pure quark fragmentation, as any established 3-quark state with strangeness. 
In contrast, if the production
mechanism of the $\tp$ state involves the fragmentation of the diquark system from the incoming proton, then the 
$\tp$ state should mainly be seen for $\eta > 0$ (towards the incoming proton direction) 
and at low transverse momenta.
In addition, high $Q^2$ DIS is more favorable~\cite{schekanov}  as in this case the incoming
proton can receive a sufficient kick from the virtual boson and the color
string can drag the diquark towards the central pseudorapidity region (small $\eta$).
Indeed, Fig.~1(right) shows that
the $1522\mev$ signal is found to occur predominantly
for $\eta > 0$. Known baryons, produced either via boson-gluon mechanism
($\Lambda_c$) or by  quark-fragmentation process ($\Lambda(1520)$), are
distributed uniformly over the whole measured pseudorapidity region $-1.5<\eta<1.5$
(see Fig.~2).

\begin{center}
\begin{minipage}[c]{0.49\textwidth}
\includegraphics[width=7.0cm,angle=0]{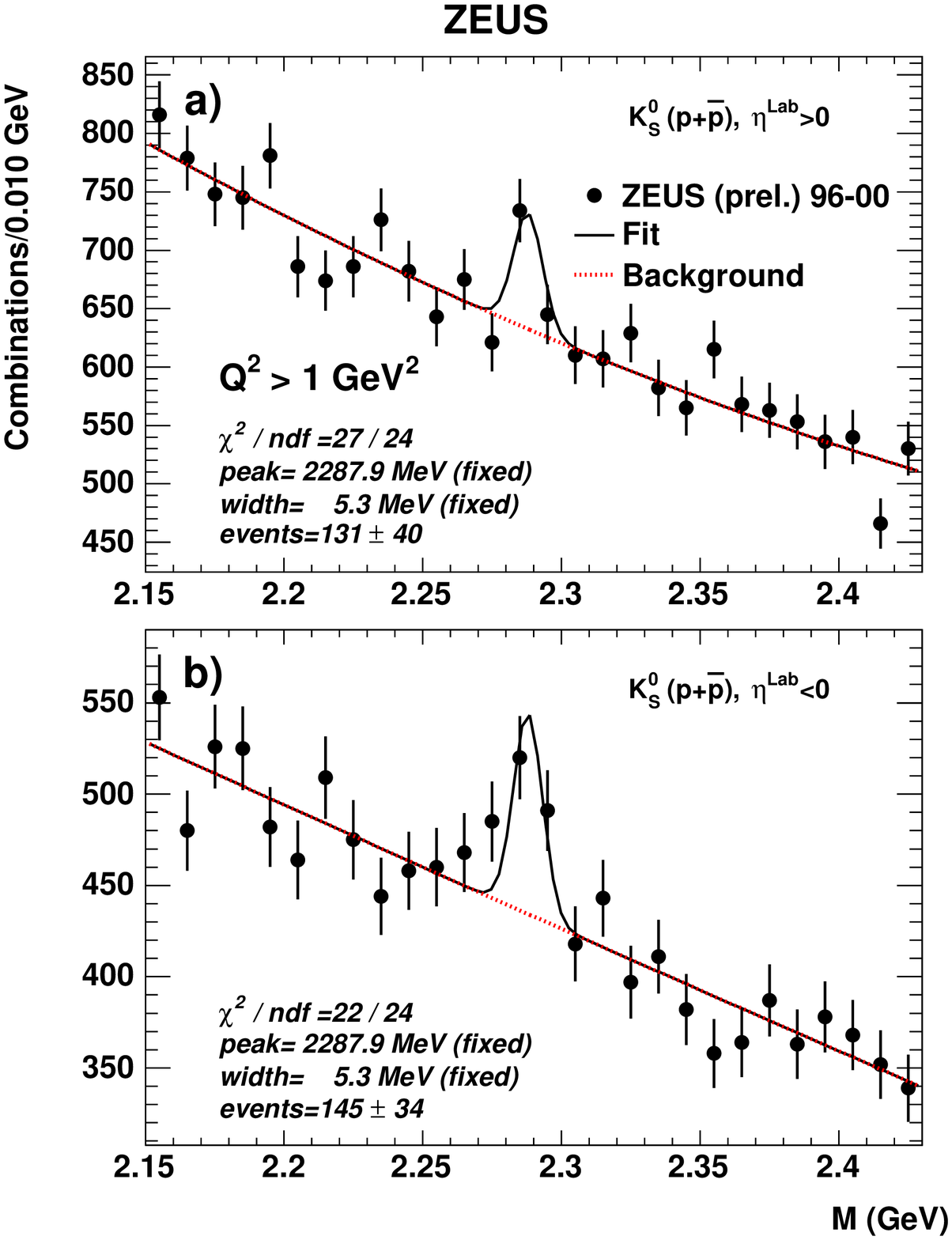}
\end{minipage}
\hfill
\begin{minipage}[c]{.49\textwidth}
\includegraphics[width=7.0cm,angle=0]{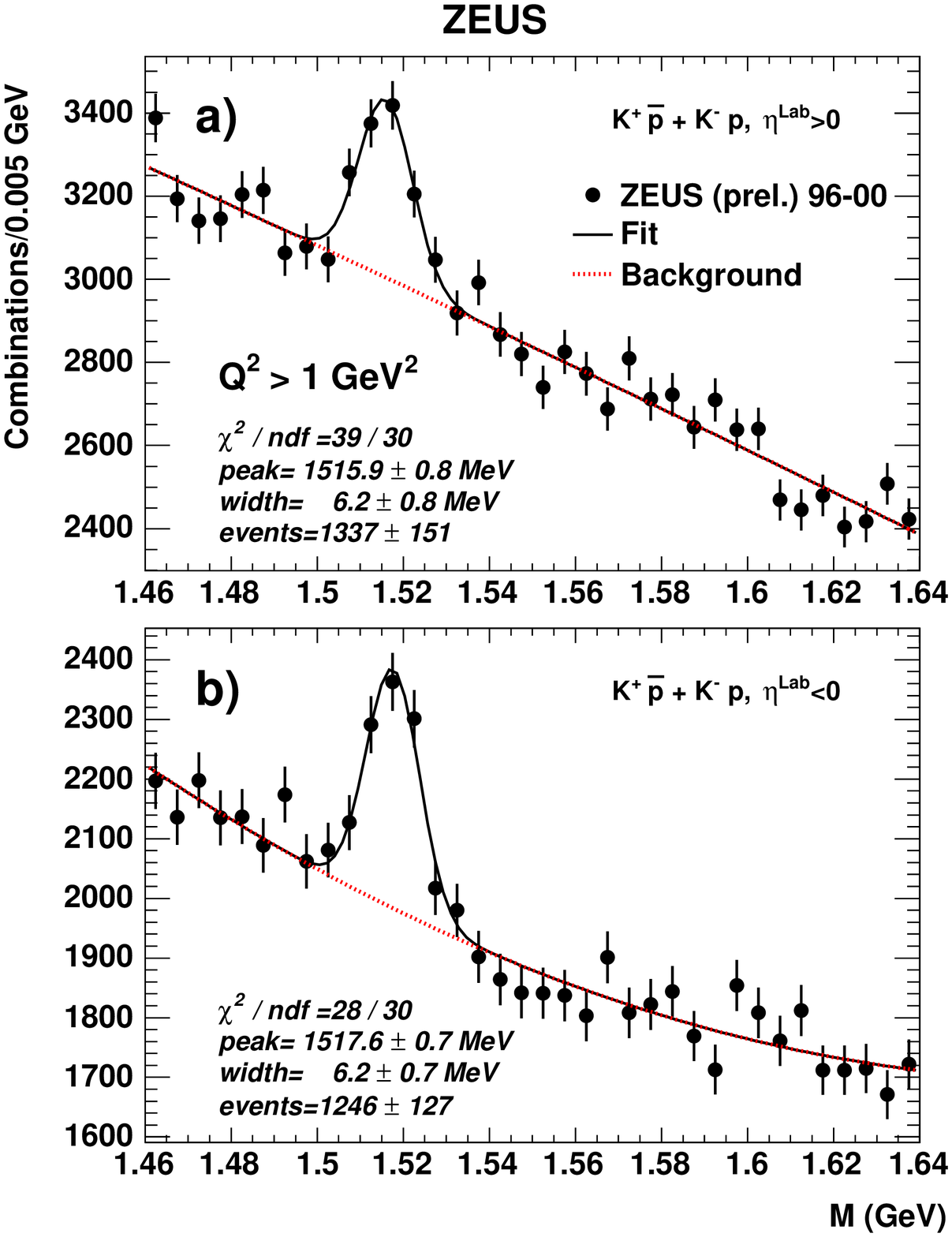}
\label{ks}
\end{minipage}
\end{center}
{\bf Figure~2.}~Left: {\it 
$\ksppb$ invariant-mass distribution near the $\Lambda_c$ mass
for  $\eta>0$ and $\eta<0$ regions for  $Q^2>1\gev^2$.
The peak position and the width were fixed from the fit to
the sum of these two mass distributions.}
Right: {\it The $K^-p(K^+\bar{p})$ invariant-mass distribution near the  $\Lambda(1520)$ mass
in  the forward ($\eta>0$) and in the rear ($\eta<0$) regions for $Q^2>1\gev^2$.
The fits  were performed using a Gaussian with a second-order polynomial function.}

\vspace{0.5cm}

\section{H1 results}

The H1 Collaboration has performed
$\tp$ searches using a similar $\ksppb$ reconstruction procedure~\cite{h1_eps05} 
as for the ZEUS analysis,
but the DIS selection was  somewhat tighter than in the ZEUS case.
An integrated luminosity of $71$ pb$^{-1}$ was used from HERA~I.
The $\ksppb$ spectrum is shown in Fig.~3(left). No signal is  observed.
Given the absence of the $\tp$  signal,
$95\%$ C.L. limits were set on the production of new states
decaying to $\ksppb$ in the mass range 1480--1700 $\mev$
in DIS for $Q^2>20\gev^2$. The limits shown in Fig.~3(right) were obtained
assuming that the production kinematics of $\tp$ is similar
to the  $\Sigma^+$ state produced by pure quark-fragmentation process.
Around the mass of 1520$\mev$, an upper limit on the cross section of roughly
100--120 pb was found, 
which is about the signal cross section 
observed\footnote{The ZEUS cross 
section~\cite{zeus_eps05b} for the $\tp$ candidates and their
antiparticles measured in the kinematic region given by $Q^2 \ge 20\gev^2$,
$0.04<y<0.95$, $p_T>0.5\gev$ and $|\eta|<1.5$ was $\sigma(e^{\pm} p \rightarrow
 e^{\pm} \>  \Theta^+\> X \rightarrow e^{\pm} \>  K^0  p\>  X) = 125  \pm 27(\mbox{stat.})^{+36}_{-28
}(\mbox{syst.}) \mbox{ pb}$} by ZEUS.

In addition, H1 investigated the $\ksppb$ invariant masses at lower $Q^2$, using an 
alternative selection
which uses  higher-momentum protons, $p>1.5\gev$. In this case,
a lower proton purity is expected. The $\ksp$ and $\kspb$ mass
combinations were analysed separately.  
As for the 
ZEUS studies~\cite{zeus}, a $\tp$ signal was not seen at low $Q^2$ and
for high-momentum protons.

\begin{center}
\begin{minipage}[c]{0.49\textwidth}
\includegraphics[width=7.0cm,angle=0]{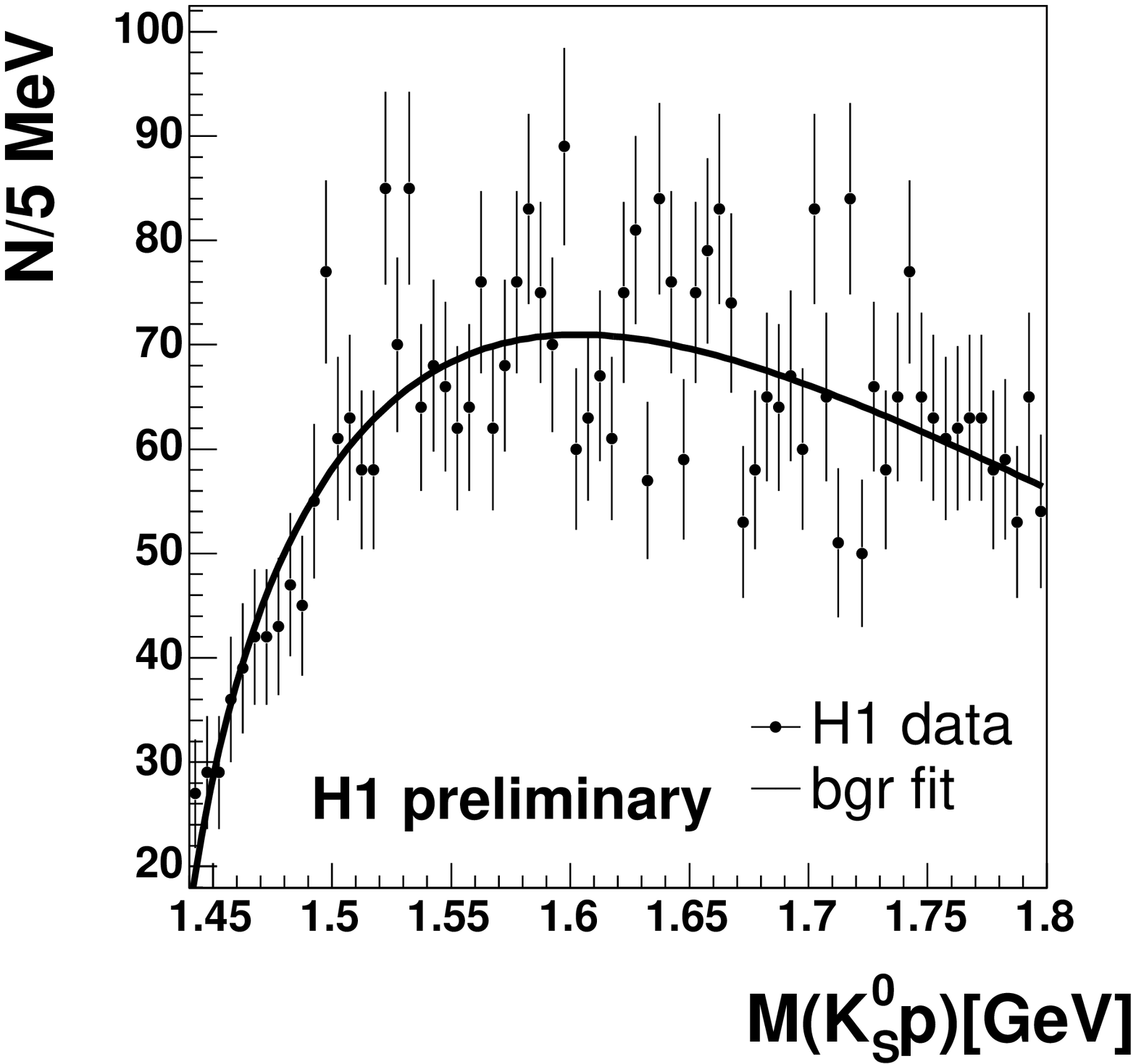}
\end{minipage}
\hfill
\begin{minipage}[c]{.49\textwidth}
\includegraphics[width=8.0cm,angle=0]{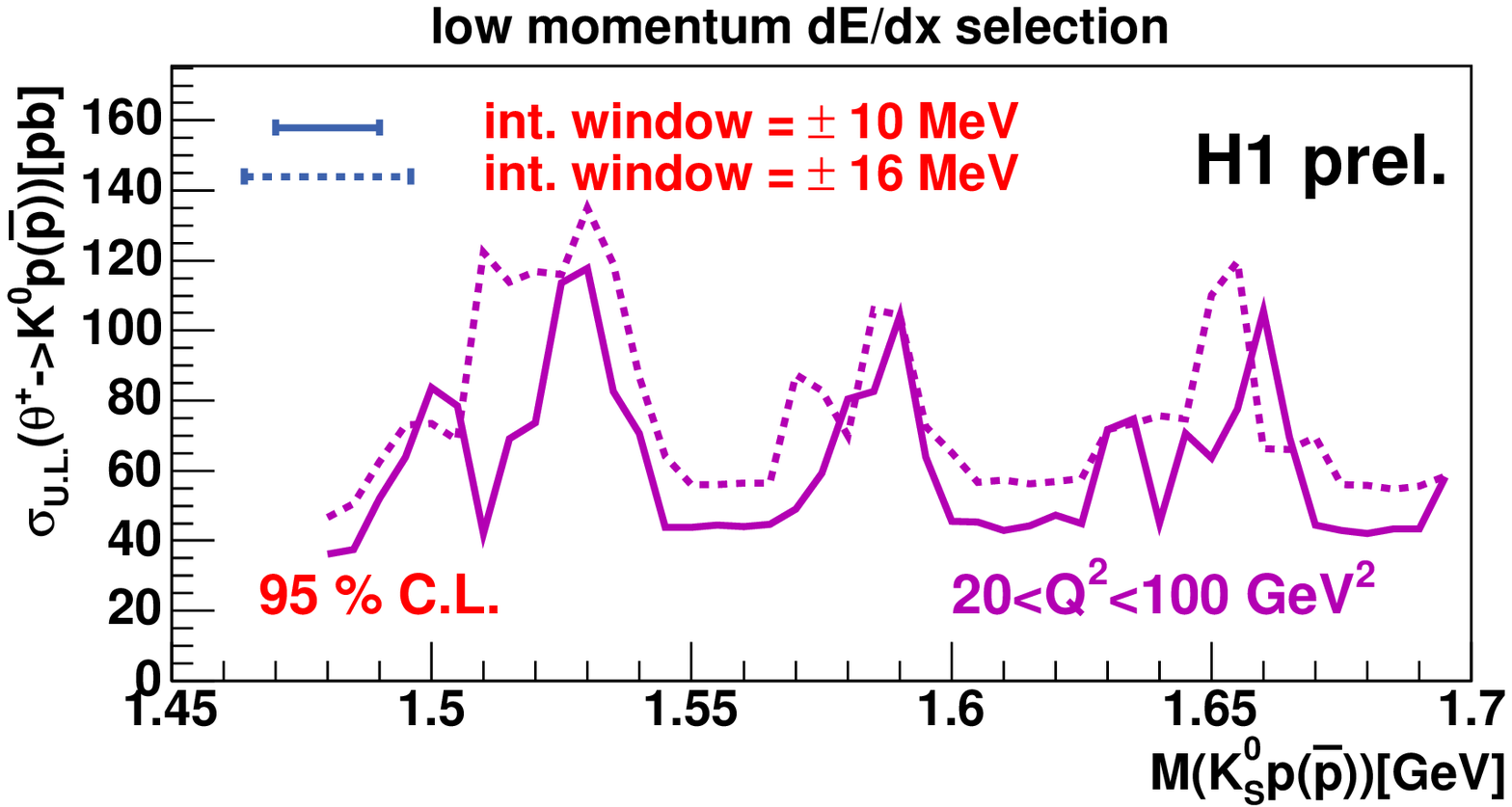}
\label{theta}
\end{minipage}
\end{center}
{\bf Figure~3.}~Left: {\it 
$\ksppb$ invariant-mass distribution near the  $1520\mev$ mass
in DIS for $Q^2>20\gev^2$ reconstructed by H1.} 
Right: {\it Upper limits on the cross section  
$\sigma(e^{\pm} p \rightarrow
e^{\pm} \>  \Theta^+\> X \rightarrow e^{\pm} \>  K^0  p\>  X)$
}


\section{$\Xi^{--}_{3/2}$  and $\Xi^{0}_{3/2}$ states?}

The $\tp$ lies at the apex of a hypothetical anti-decuplet of pentaquarks 
with spin $1/2$ \cite{zp:a359:305}. The baryonic states
$\Xi^{--}_{3/2}$ and $\Xi^{0}_{3/2}$ at the bottom of this
antidecuplet  are also manifestly exotic. A support for this
picture came from the NA49 experiment which recently made
observations~\cite{prl92:042003}  of both states near $1862\mev$ in the 
$\xp$ and $\xbp$ decay channels in fixed-target $pp$
collisions at the CERN SPS.

ZEUS has performed a similar analysis~\cite{zeus-xi}  by
combining $\Lambda$ with $\pi$
using displaced tertiary vertices.
With more than $190$ reconstructed $\Xi^0(1530)$ near the mass threshold of
$\Xi\pi$ spectrum, no pentaquark signal was observed near the $1860\mev$ mass region.
The $95\%$ C.L. upper limit on the ratio $\Xi^{--}_{3/2}(\Xi^0_{3/2})$ to $\Xi^0(1530)$
was in the range $0.2-0.45$.

\end{document}